\documentclass[11pt]{article}          
\usepackage{epsf,mathrsfs}
\makeatletter

\@addtoreset{equation}{section}
\makeatother

\def\omit#1{_{\!\rlap{$\scriptscriptstyle \backslash$}
{\scriptscriptstyle #1}}}

\title{\centerline{\normalsize SINP/TNP/02-26 \hfill hep-ph/0209053}
\bf Inverse beta-decay of arbitrarily polarized neutrons in a
magnetic field} 

\author{
{\bf Kaushik Bhattacharya} and 
{\bf Palash B. Pal\thanks{e-mail addresses: kaushikb@theory.saha.ernet.in,
pbpal@theory.saha.ernet.in}}\\ 
\normalsize Saha Institute of Nuclear Physics, 1/AF Bidhan-Nagar, 
Calcutta 700064, India}

\date{}

\begin{document}

\maketitle 

\begin{abstract} \noindent\small
We calculate the cross section of the inverse beta decay process,
$\nu_e+n\to p+e$, in a background magnetic field which is much smaller
than $m_p^2/e$.  Using exact solutions of the Dirac equation in a
constant magnetic field, we find the cross section for arbitrary
polarization of the initial neutrons.  The cross section depends on
the direction of the incident neutrino even when the initial neutron
is assumed to be at rest and has no net polarization.  Possible
implications of the result are discussed.

\end{abstract} 

\section{Introduction}\label{in} 
The interactions of elementary particles show novel features when they
occur in non-trivial backgrounds.  Study of particle propagation in
matter has proved pivotal in the understanding of the solar neutrino
problem.  Similar studies of particle processes in background magnetic
fields are also important since stellar objects like neutron stars are
expected to possess very high magnetic fields, of the order of
$10^{12}$\,G or higher.  Analysis of these processes might be crucial
for obtaining a proper understanding of the properties of these stars.

In this paper, we calculate the cross section of the inverse
beta-decay process in a magnetic field.  We consider the possibility
that the neutrons may be totally or partially polarized in the
magnetic field, and find the cross section as a function of this
polarization.  The neutrinos are assumed to be strictly standard model
neutrinos, without any mass and consequent properties.  The presence
of the magnetic field breaks the isotropy of the background, and a
careful calculation in this background reveals a dependence of the
cross section on the incident neutrino direction with respect to the
magnetic field.

Considerable work has been done on the magnetic field dependence of
the URCA processes which have neutrinos in their final states
\cite{DRT, MOc, Gvozdev:1999md, Arras:1998mv, Baiko:1998jq}.  An angular dependence
obtained in the differential cross section of these reactions imply
that in a star with high magnetic field, neutrinos are created
asymmetrically with respect to the magnetic field direction.  The
process that we consider, on the other hand, have neutrinos in the
initial state. So this process influences the neutrino opacity in a
star.

Some calculations of this process exist in the literature.  Roulet
\cite{Roulet:1997sw}, as well as Lai and Qian \cite{Lai:1998sz}
performed the calculation by assuming that the magnetic field effects
enter only through the phase space integrals, whereas the matrix
element remains unaffected.  Gvozdev and Ognev \cite{Gvozdev:1999md}
considered the final electron to be exclusively in the lowest Landau
level.  Arras and Lai \cite{Arras:1998mv} calculated only the angular
asymmetry, and only to the first order in the background magnetic
field.  The earlier calculations of the present authors
\cite{Bhattacharya:1999bm,KP2} did not take neutron polarization into
account.  In this paper, we consider the problem in full detail ---
i.e., we calculate the matrix element using spinor solutions of the
electron in a magnetic field, take all possible final Landau levels
into account, include the possiblity of neutron polarization, and
perform the calculations to all orders in the background field in the
4-fermi interaction theory.

The paper is organized as follows. In Sec~\ref{so}, we provide some
background for the calculation. Most of this section contains no
original material, but we provide it for the sake of completeness, as
well as for setting up the notation that will be used in the later
sections. In Sec.~\ref{fo}, we define the fermion field operator and
show how it acts on the states in the presence of a magnetic
field. Sec.~\ref{ib} contains the calculation of the cross section for
a monochromatic neutrino beam, which contains the main results of our
paper. In Sec.~\ref{es}, we discuss the realistic case where the
initial neutrino beam has a finite energy spread.  Sec.~\ref{co}
contains our discussions and conclusions.

\section{Solutions of the Dirac equation in a uniform magnetic
field}\label{so}
For a particle of charge $eQ$, the Dirac equation in presence of a
magnetic field is given by 
\begin{eqnarray}
i {\partial\psi \over \partial t} = \big[ \vec \alpha \cdot
(-i\vec\nabla - eQ\vec A) + \beta m \big] \psi \,,
\label{DiracEq}
\end{eqnarray}
where $\vec\alpha$ and $\beta$ are the Dirac matrices, and $\vec A$ is
the vector potential.  In our convention, $e$ is the positive unit of
charge, taken as usual to be equal to the proton charge.

For stationary states, we can write
\begin{eqnarray}
\psi = e^{-iEt} \left( \begin{array}{c} \phi \\ \chi \end{array}
\right) \,,
\end{eqnarray}
where $\phi$ and $\chi$ are 2-component objects.  We use the Pauli-Dirac
representation of the Dirac matrices, in which
\begin{eqnarray}
\vec \alpha = \left( \begin{array}{cc} 
		0 & \vec\sigma \\
		\vec\sigma & 0
	      \end{array} \right) \,, \qquad
\beta = \left( \begin{array}{cc} 
		1 & 0 \\
		0 & 1 
	      \end{array} \right)
\end{eqnarray}
where each block represents a $2\times2$ matrix, and $\vec\sigma$ are
the Pauli matrices.  With this notation, we can write Eq.\
(\ref{DiracEq}) as
\begin{eqnarray}
(E-m)\phi &=& \vec \sigma \cdot (-i\vec\nabla - eQ\vec A) \chi \,, 
\label{eq1}\\*
(E+m) \chi &=& \vec \sigma \cdot (-i\vec\nabla - eQ\vec A) \phi \,.
\label{eq2}
\end{eqnarray}
Eliminating $\chi$, we obtain 
\begin{eqnarray}
(E^2 - m^2)\phi &=& \Big[ \vec \sigma \cdot (-i\vec\nabla - eQ\vec A) 
\Big]^2 \phi \,.
\label{phieq1} 
\end{eqnarray}
We will work with a constant magnetic field $\vec B$.  Without loss of
generality, it can be taken along the $z$-direction. The vector
potential can be chosen in many equivalent ways. We take
\begin{eqnarray}
A_0 = A_y = A_z = 0 \,, \qquad A_x = -yB\,.
\label{A}
\end{eqnarray}
With this choice, Eq.\ (\ref{phieq1}) reduces to the form
\begin{eqnarray}
(E^2 - m^2)\phi 
&=& \Big[ -\vec\nabla^2 + (eQB)^2 y^2 - eQB(2iy {\partial\over
\partial x} + \sigma_z) \Big] \phi \,. 
\label{phieq} 
\end{eqnarray}
Noticing that the co-ordinates $x$ and $z$ do not appear in the
equation except through the derivatives, we can write the
solutions as
\begin{eqnarray}
\phi = e^{i \vec {\scriptstyle p} \cdot \vec{\scriptstyle X} \omit y}
f(y) \,, 
\label{phiform}
\end{eqnarray}
where $f(y)$ is a 2-component matrix which depends only on the
$y$-coordinate, and possibly some momentum components, as we will see
shortly. We have also introduced the notation $\vec X$ for the spatial
co-ordinates (in order to distinguish it from $x$, which is one of the
components of $\vec X$), and $\vec X\omit y$ for the vector $\vec X$
with its $y$-component set equal to zero. In other words, $\vec p\cdot
\vec X{\omit y} \equiv p_xx+p_zz$, where $p_x$ and $p_z$ denote the
eigenvalues of momentum in the $x$ and $z$ directions.\footnote{It is
to be understood that whenever we write the spatial component of any
vector with a lettered subscript, it would imply the corresponding
contravariant component of the relevant 4-vector.}

There will be two independent solutions for $f(y)$, which can be
taken, without any loss of generality, to be the eigenstates of
$\sigma_z$ with eigenvalues $s=\pm 1$. This means that we choose the
two independent solutions in the form
\begin{eqnarray}
f_+ (y) = \left( \begin{array}{c} F_+(y) \\ 0 \end{array} \right) \,,
\qquad 
f_- (y) = \left( \begin{array}{c} 0 \\ F_-(y) \end{array} \right) \,.
\end{eqnarray}
Since $\sigma_z F_s = sF_s$, the differential equations satisfied by
$F_s$ is
\begin{eqnarray}
{d^2F_s \over dy^2} - (eQBy + p_x)^2 F_s + (E^2 - m^2 - p_z^2 + eQBs)
F_s = 0 \,,
\label{Fseqn}
\end{eqnarray}
which is obtained from Eq.\ (\ref{phieq}).  The solution is obtained
by using the dimensionless variable
\begin{eqnarray}
\xi = \sqrt{e |Q| B} \left( y + {p_x \over eQB} \right) \,,
\label{xi}
\end{eqnarray}
which transforms Eq.\ (\ref{Fseqn}) to the form
\begin{eqnarray}
\left[ {d^2 \over d\xi^2} -\xi^2 + a_s \right] F_s = 0 \,,
\end{eqnarray}
where
\begin{eqnarray}
a_s = {E^2 - m^2 - p_z^2 + eQBs \over e|Q|B} \,.
\end{eqnarray}
This is a special form of Hermite's equation, and the solutions exist
provided $a_s=2\nu+1$ for $\nu=0,1,2,\cdots$. This provides the energy
eigenvalues 
\begin{eqnarray}
E^2 = m^2 + p_z^2 + (2\nu+1)e|Q|B - eQBs \,,
\end{eqnarray}
and the solutions for $F_s$ are
\begin{eqnarray}
N_{\nu} e^{-\xi^2/2} H_{\nu}(\xi) \equiv I_{\nu}(\xi) \,,
\label{In}
\end{eqnarray}
where $H_\nu$ are Hermite polynomials of order $\nu$, and $N_\nu$ are
normalizations which we take to be
\begin{eqnarray}
N_\nu = 
\left( {\sqrt{e|Q|B} \over \nu! \, 2^\nu \sqrt{\pi}} \, \right)^{1/2} \,. 
\label{Nn}
\end{eqnarray}
We stress that the choice of normalization can be arbitrarily made, as
will be clarified later.  With our choice, the functions $I_\nu$
satisfy the completeness relation
\begin{eqnarray}
\sum_\nu I_\nu(\xi) I_\nu(\xi_\star) = \sqrt{e|Q|B} \;
\delta(\xi-\xi_\star) = \delta (y-y_\star) \,, 
\label{completeness}
\end{eqnarray}
where $\xi_\star$ is obtained by replacing $y$ by $y_\star$ in Eq.\
(\ref{xi}).

So far, $Q$ was arbitrary.  We now specialize to the case of
electrons, for which $Q=-1$.  The solutions are then conveniently
classified by the energy eigenvalues
\begin{eqnarray}
E_n^2 = m^2 + p_z^2 + 2neB \,,
\label{En}
\end{eqnarray}
which is the relativistic form of Landau energy levels. The solutions
are two fold degenerate in general: for $s=1$, $\nu=n-1$ and for
$s=-1$, $\nu=n$.  In the case of $n=0$, only the second solution is
available since $\nu$ cannot be negative.  The solutions can have
positive or negative energies. We will denote the positive square root
of the right side by $E_n$. Representing the solution corresponding to
this $n$-th Landau level by a superscript $n$, we can then write for
the positive energy solutions,
\begin{eqnarray}
f_+^{(n)} (y) = \left( \begin{array}{c} 
I_{n-1}(\xi) \\ 0 \end{array} \right) \,,
\qquad 
f_-^{(n)} (y) = \left( \begin{array}{c} 
0 \\ I_n (\xi) \end{array} \right) \,.
\label{fsolns}
\end{eqnarray}
For $n=0$, the solution $f_+$ does not exist. We will consistently
incorporate this fact by defining
\begin{eqnarray}
I_{-1} (y) = 0 \,,
\label{I_-1}
\end{eqnarray}
in addition to the definition of $I_n$ in Eq.\ (\ref{In}) for
non-negative integers $n$.

The solutions in Eq.\ (\ref{fsolns}) determine the upper components of
the spinors through Eq.\ (\ref{phiform}). The lower
components, denoted by $\chi$ earlier, can be solved using
Eq.\ (\ref{eq2}), and finally the positive energy solutions of the
Dirac equation can be written as
\begin{eqnarray}
e^{-ip\cdot X {\omit y}} U_s (y,n,\vec p \omit y) \,,
\end{eqnarray}
where $X^\mu$ denotes the space-time coordinate. And $U_s$ are given
by
\begin{eqnarray}
U_+ (y,n,\vec p \omit y) = \left( \begin{array}{c} 
I_{n-1}(\xi) \\[2ex] 0 \\[2ex] 
{\strut\textstyle p_z \over \strut\textstyle E_n+m} I_{n-1}(\xi) \\[2ex]
-\, {\strut\textstyle \sqrt{2neB} \over \strut\textstyle 
E_n+m} I_n (\xi) 
\end{array} \right) \,, \qquad 
U_- (y,n,\vec p \omit y) = \left( \begin{array}{c} 
0 \\[2ex] I_n (\xi) \\[2ex]
-\, {\strut\textstyle \sqrt{2neB} \over \strut\textstyle E_n+m}
I_{n-1}(\xi) \\[2ex] 
-\,{\strut\textstyle p_z \over \strut\textstyle E_n+m} I_n(\xi) 
\end{array} \right) \,. 
\label{Usoln}
\end{eqnarray}

A similar procedure can be adopted for negative energy spinors which
have energy eigenvalues $E=-E_n$. In this case, it is easier to start
with the two lower components first and then find the upper components
from Eq.\ (\ref{eq1}). The solutions are
\begin{eqnarray}
e^{ip\cdot X{\omit y}} V_s (y,n, \vec p\omit y) \,,
\end{eqnarray}
where 
\begin{eqnarray}
V_+ (y,n,\vec p\omit y) = \left( \begin{array}{c} 
{\strut\textstyle p_z \over \strut\textstyle E_n+m}
I_{n-1}(\widetilde\xi) \\[2ex] 
{\strut\textstyle \sqrt{2neB} \over \strut\textstyle E_n+m} 
I_n (\widetilde\xi)  \\[2ex] 
I_{n-1}(\widetilde\xi) \\[2ex] 0
\end{array} \right) \,, \qquad 
V_- (y,n,\vec p\omit y) = \left( \begin{array}{c} 
{\strut\textstyle \sqrt{2neB} \over \strut\textstyle E_n+m}
I_{n-1}(\widetilde\xi) \\[2ex] 
-\,{\strut\textstyle p_z \over \strut\textstyle E_n+m}
I_n(\widetilde\xi)  \\[2ex] 
0 \\[2ex] I_n (\widetilde\xi)
\end{array} \right) \,.
\label{Vsoln}
\end{eqnarray}
where $\widetilde\xi$ is obtained from $\xi$ by changing the sign of
the $p_x$-term.

For future use, we note down a few identities involving the spinors
which can be obtained by direct substitutions of the solutions
obtained above. The spin sum for the $U$-spinors is
\begin{eqnarray}
P_U (y,y_\star ,n,\vec p\omit y) &\equiv&
\sum_s U_s (y,n,\vec p\omit y) \overline U_s (y_\star ,n,\vec p\omit
y) \nonumber\\*  
& = &
{1\over 2(E_n+m)} \times 
\begin{array}[t]{l}
\bigg[ \left\{ m(1+\sigma_z) +
\rlap/p_\parallel - 
\widetilde{\rlap/p}_\parallel \gamma_5 \right\} I_{n-1}(\xi)
I_{n-1}(\xi_\star) \\ 
+ \left\{ m(1-\sigma_z) + \rlap/p_\parallel +
\widetilde{\rlap/p}_\parallel \gamma_5 \right\} I_n(\xi)
I_n (\xi_\star) \\ 
- \sqrt{2neB} (\gamma_1 - i\gamma_2) I_n(\xi) I_{n-1}(\xi_\star) \\
- \sqrt{2neB} (\gamma_1 + i\gamma_2) I_{n-1}(\xi) I_n(\xi_\star) 
\bigg] \,,
\end{array}
\nonumber\\* 
\label{PU}
\end{eqnarray}
where we have introduced the following notations for any object $a$
carrying a Lorentz index:
\begin{eqnarray}
a_\parallel^\mu &=& (a_0, 0, 0, a_z) \,, \nonumber\\
\widetilde a_\parallel^\mu &=& (a_z, 0, 0, a_0) \,. 
\end{eqnarray}
Similarly, the spin sum for the $V$-spinors can also be calculated,
and we obtain
\begin{eqnarray}
P_V (y,y_\star ,n,\vec p\omit y) &\equiv&
\sum_s V_s (y,n,\vec p\omit y) \overline V_s (y,n,\vec p\omit y) 
\nonumber\\* 
& = &
{1\over 2(E_n+m)} \times 
\begin{array}[t]{l}
\Bigg[ \left\{ -m(1+\sigma_z) +
\rlap/p_\parallel - 
\widetilde{\rlap/p}_\parallel \gamma_5 \right\} I_{n-1}(\widetilde\xi)
I_{n-1} (\widetilde\xi _\star) \\ 
+ \left\{ -m(1-\sigma_z) + \rlap/p_\parallel +
\widetilde{\rlap/p}_\parallel \gamma_5 \right\} I_n(\widetilde\xi)
I_n(\widetilde\xi _\star) \\ 
+ \sqrt{2neB} (\gamma_1 - i\gamma_2) I_n(\widetilde\xi)
I_{n-1}(\widetilde\xi _\star) \\ 
+ \sqrt{2neB} (\gamma_1 + i\gamma_2) I_{n-1}(\widetilde\xi)
I_n(\widetilde\xi 
_\star) \Bigg] \,. 
\end{array} \nonumber\\*
\end{eqnarray}
%

\section{The fermion field operator}\label{fo}
Since we have found the solutions to the Dirac equation, we can now
use them to construct the fermion field operator in the second
quantized version. For this, we write
\begin{eqnarray}
\psi(X) = \sum_{s=\pm} \sum_{n=0}^\infty \int {dp_x \, dp_z \over D}
\left[ f_s (n,\vec p\omit y) e^{-ip\cdot X {\omit y}} U_s (y,n,\vec p\omit y) + 
\widehat f_s^\dagger (n,\vec p\omit y) e^{ip\cdot X {\omit y}} V_s
(y,n,\vec p\omit y) \right] \,.
\label{2ndquant}
\end{eqnarray}
Here, $f_s(n,\vec p\omit y)$ is the annihilation operator for the
fermion, and $\widehat f_s^\dagger(n,\vec p\omit y)$ is the creation
operator for the antifermion in the $n$-th Landau level with given
values of $p_x$ and $p_z$. The creation and annihilation operators
satisfy the anticommutation relations
\begin{eqnarray}
\left[ f_s (n,\vec p\omit y), f_{s'}^\dagger (n',\vec p'\omit y)
\right]_+ = 
\delta_{ss'} \delta_{nn'} \delta(p_x-p'_x) \delta (p_z - p'_z) \,,
\label{freln}
\end{eqnarray}
and a similar one with the operators $\widehat f$ and $\widehat
f^\dagger$, all other anticommutators being zero. The quantity $D$
appearing in Eq.\ (\ref{2ndquant}) depends on the normalization of the
spinor solutions, and this is why of the spinors could have been
chosen arbitrarily, as remarked after Eq.\ (\ref{Nn}).  Once we have
chosen the spinor normalization, the factor $D$ appearing in Eq.\
(\ref{2ndquant}) is however fixed, and it can be determined from the
equal time anticommutation relation
\begin{eqnarray}
\left[ \psi(X), \psi^\dagger(X_\star) \right]_+ = \delta^3 (\vec X - \vec
X_\star) \,.
\label{anticomm}
\end{eqnarray}
Plugging in the expression given in
Eq.\ (\ref{2ndquant}) to the left side of this equation and using the
anticommutation relations of Eq.\ (\ref{freln}), we obtain
\begin{eqnarray}
\left[ \psi(X), \psi^\dagger(X_\star) \right]_+ = \sum_{s} \sum_{n} \int
{dp_x \, dp_z \over D^2} && 
\Big( e^{-ip_x(x-x_\star)} e^{-ip_z(z-z_\star)} 
U_s (y,n,\vec p\omit y) U_s^\dagger (y_\star ,n,\vec p\omit y)
\nonumber\\* 
&& +  e^{ip_x(x-x_\star)} e^{ip_z(z-z_\star)} 
V_s (y,n,\vec p\omit y) V_s^\dagger (y_\star ,n,\vec p\omit y) \Big) \,.
\end{eqnarray}
Changing the signs of the dummy integration variables $p_x$ and $p_z$
in the second term, we can rewrite it as
\begin{eqnarray}
\left[ \psi(X), \psi^\dagger(X_\star) \right]_+ = \sum_{s} \sum_{n} \int
{dp_x \, dp_z \over D^2} && e^{-ip_x(x-x_\star)}
e^{-ip_z(z-z_\star)} \Big( 
U_s (y,n,\vec p\omit y) U_s^\dagger (y_\star ,n,\vec p\omit y)
\nonumber\\* 
&& +  V_s (y,n,-\vec p\omit y) V_s^\dagger (y_\star ,n,-\vec p\omit y)
\Big) \,. 
\label{anticomm1}
\end{eqnarray}
Using now the solutions for the $U$ and the $V$ spinors from
Eqs. (\ref{Usoln}) and (\ref{Vsoln}), it is straight forward to verify
that 
\begin{eqnarray}
&& \sum_s \Big( U_s (y,n,\vec p\omit y) U_s^\dagger (y_\star ,n,\vec
p\omit y) 
+  V_s (y,n,-\vec p\omit y) V_s^\dagger (y_\star ,n,-\vec p\omit y) \Big)
\nonumber\\* 
&=& \left( 1 + {p_z^2 + 2neB \over (E_n+m)^2} \right) \times {\rm
diag} \; \Big [ I_{n-1}(\xi) I_{n-1}(\xi_\star), 
I_n(\xi) I_n(\xi_\star),
I_{n-1}(\xi) I_{n-1}(\xi_\star),  I_n(\xi) I_n(\xi_\star) \Big] \,,
\label{ssum}
\end{eqnarray}
where `diag' indicates a diagonal matrix with the specified entries,
and $\xi$ and $\xi_\star$ involve the same value of $p_x$. At this stage,
we can perform the sum over $n$ in Eq.\ (\ref{anticomm1}) using the
completeness relation of Eq.\ (\ref{completeness}), which gives the
$\delta$-function of the $y$-coordinate that should appear in the
anticommutator.  Finally, performing the integrations over $p_x$ and
$p_z$, we can recover the $\delta$-functions for the other two
coordinates as well, provided
\begin{eqnarray}
{2E_n \over E_n+m} \; {1\over D^2} = {1\over (2\pi)^2} \,,
\end{eqnarray}
using the expression for the energy eigenvalues from Eq.\ (\ref{En}) to
rewrite the prefactor appearing on the right side of
Eq.\ (\ref{ssum}). Putting the solution for $D$, we can rewrite
Eq.\ (\ref{2ndquant}) as
\begin{eqnarray}
\psi(X) &=& \sum_{s=\pm} \sum_{n=0}^\infty \int {dp_x \, dp_z \over
2\pi} \sqrt {E_n+m \over 2E_n} \nonumber\\* && \times
\left[ f_s (n,\vec p\omit y) e^{-ip\cdot X {\omit y}} U_s (y,n,\vec p\omit y) + 
\widehat f_s^\dagger (n,\vec p\omit y) e^{ip\cdot X {\omit y}} V_s
(y,n,\vec p\omit y) \right] \,.
\label{psi}
\end{eqnarray}

The one-fermion states are defined as
\begin{eqnarray}
\left| n,\vec p\omit y \right> = C f^\dagger (n,\vec p\omit y) \left|
0 \right> \,. 
\end{eqnarray}
The normalization constant $C$ is determined by the condition that the
one-particle states should be orthonormal. For this, we need to define
the theory in a finite but large region whose dimensions are  $L_x$,
$L_y$ and $L_z$ along the three spatial axes. This gives
\begin{eqnarray}
C = {2\pi \over \sqrt{L_x L_z}} \,.
\end{eqnarray}
Then
\begin{eqnarray}
\psi_U(X) \left| n,\vec p\omit y \right> = \sqrt {E_n+m \over 2E_n L_xL_z}
e^{-ip\cdot X {\omit y}} U_s (y,n,\vec p\omit y) \left| 0 \right> \,,
\label{psiket}
\end{eqnarray}
where $\psi_U$ denotes the term in Eq.\ (\ref{psi}) that contains the
$U$-spinors. Similarly, 
\begin{eqnarray}
\left< n,\vec p\omit y \left| \overline \psi_U(X) \right. \right.
= \sqrt {E_n+m \over 2E_n L_xL_z}
e^{ip\cdot X \omit y} \overline U_s (y,n,\vec p\omit y) \left< 0
\right| \,.
\label{brapsi}
\end{eqnarray}
%

\section{Inverse beta-decay}\label{ib} 
In this section, we calculate the cross section for the inverse
beta-decay process $\nu_e+n\to p+e^-$ in a background magnetic field.
The magnetic field might provide a net polarization of the neutrons,
which we take into account. However, the magnitude of the field is
assumed to be much smaller than $m_n^2/e$ or $m_p^2/e$, so we ignore
its effects on the proton and neutron spinors.  The electron spinors,
on the other hand, are the ones appropriate for the Landau
levels. Thus, we can write the process as
\begin{eqnarray}
\nu_e(\vec k) + n(\vec P) \to p(\vec P') + e(\vec p'\omit y, n') \,.
\label{invbeta}
\end{eqnarray}
%

\subsection{The $S$-matrix element}
The interaction Lagrangian for this process is
\begin{eqnarray}
\mathscr L_{\rm int} = \sqrt 2 \,G_\beta \left[ \overline
\psi_{(e)} \gamma^\mu L \psi_{(\nu_e)} \right] \; 
\left[ \overline
\psi_{(p)} \gamma_\mu (g_V - g_A \gamma_5) \psi_{(n)} \right] \,,
\end{eqnarray}
where $L=\frac12(1-\gamma_5)$ and $G_\beta=G_F\cos\theta_c$,
$\theta_c$ being the Cabibbo angle. In first order perturbation, the
$S$-matrix element between the final and the initial states of the
process in Eq.\ (\ref{invbeta}) is therefore given by
\begin{eqnarray}
S_{fi} = \sqrt 2 \, G_\beta \int d^4X &&
\left< e(\vec p'\omit y, n') \left| \overline
\psi_{(e)} \gamma^\mu L \psi_{(\nu_e)} 
\right| \nu_e (\vec k) \right> \nonumber\\*
&& \times  
\left< p(P') \left| \overline
\psi_{(p)} \gamma_\mu (g_V - g_A \gamma_5) \psi_{(n)}  \right| n(P)
\right> \,. 
\label{Sfi1}
\end{eqnarray}
For the hadronic part, we should use the usual solutions of the Dirac
field which are normalized within a box of volume $V$, and this gives
\begin{eqnarray}
\left< p(P') \left| \overline
\psi_{(p)} \gamma_\mu (g_V - g_A \gamma_5) \psi_{(n)}  \right| n(P)
\right> 
&=& {e^{i(P'-P)\cdot X} \over \sqrt{2{\cal E} V}
\sqrt{2{\cal E}' V}} \; 
\left[ \overline u_{(p)}(\vec P') \gamma_\mu (g_V - g_A \gamma_5)
u_{(n)}(\vec P) \right], \quad
\end{eqnarray}
using the notations ${\cal E}=P_0$ and ${\cal E}'=P'_0$. For the
leptonic part, we need to take into account the magnetic spinors for
the electron. Using Eq.\ (\ref{brapsi}), we obtain
\begin{eqnarray}
\left< e(\vec p'\omit y, n') \left| \overline
\psi_{(e)} \gamma^\mu L \psi_{(\nu_e)} 
\right| \nu_e (\vec k) \right> 
&=& {e^{-ik\cdot X + ip'\cdot X\omit y} \over \sqrt{2\omega V}}
\sqrt{E_{n'}+m \over 
2E_{n'}L_xL_z} 
\left[ \overline U_{(e)}(y,n',\vec p'\omit y) \gamma^\mu L
u_{(\nu_e)}(\vec k) \right]. \quad
\end{eqnarray}
Putting these back into Eq.\ (\ref{Sfi1}) and performing the
integrations over all co-ordinates except $y$, we obtain
\begin{eqnarray}
S_{fi} &=& (2\pi)^3 \delta^3 \omit y (P+k-P'-p')
\left[E_{n'}+m \over 2\omega V \; 2{\cal E}V \; 2{\cal E}'V
2E_{n'}L_xL_z \right]^{1/2} {\mathscr M}_{fi} \,.
\label{Sfi2}
\end{eqnarray}
Here, $\delta^3\omit y$ implies, in accordance with the notation
introduced earlier, the $\delta$-function for all space-time
co-ordinates except $y$. Contrary to the field-free case, we do not
get 4-momentum conservation because the $y$-component of momentum is
not a good quantum number in this problem. The quantity ${\mathscr
M}_{fi}$ is the Feynman amplitude, given by
\begin{eqnarray}
{\mathscr M}_{fi} = \surd 2 G_\beta
\Big[ \overline u_{(p)}(\vec P') \gamma_\mu (g_V - g_A\gamma_5)
u_{(n)}(\vec P) \Big]  
\int dy \; e^{iq_yy}
\Big[ \overline U_{(e)} (y,n',\vec p'\omit y) \gamma^\mu L
u_{(\nu_e)}(\vec k) \Big] \,,
\end{eqnarray}
using the shorthand
\begin{eqnarray}
q_y = P_y+k_y-P'_y \,.
\end{eqnarray}

The transition rate in a large time $T$ is given by
$|S_{fi}|^2/T$. {}From Eq.\ (\ref{Sfi2}), using the usual rules like
\begin{eqnarray}
\Big| \delta ({\cal E}+\omega-{\cal E}' - E_{n'}) \Big|^2 
&=& {T\over 2\pi} \;
\delta ({\cal E}+\omega-{\cal E}' - E_{n'})  \,,\nonumber\\*
\Big| \delta (P_x+k_x-P'_x-p'_x) \Big|^2 &=& {L_x\over 2\pi} \;
\delta (P_x+k_x-P'_x-p'_x) \,,\nonumber\\*
\Big| \delta (P_z+k_z-P'_z-p'_z) \Big|^2 &=& {L_z\over 2\pi} \;
\delta (P_z+k_z-P'_z-p'_z) \,,
\end{eqnarray}
we obtain
\begin{eqnarray}
|S_{fi}|^2/T &=& {1\over 16} (2\pi)^3 \delta^3 \omit y (P+k-P'-p')
{E_{n'}+m \over V^3 \omega{\cal EE}' E_{n'}}
\Big| {\mathscr M}_{fi} \Big|^2 \,.
\end{eqnarray}
%

\subsection{The scattering cross section}
Using unit flux $1/V$ for the incident particle as usual, we can write
the differential cross section as
\begin{eqnarray}
d\sigma = V\, {|S_{fi}|^2\over T}d\rho \,,
\end{eqnarray}
where $d\rho$, the differential phase space for final particles, is
given in our case by
\begin{eqnarray}
d\rho = {L_x\over 2\pi} \, dp'_x \; {L_z\over 2\pi} \, dp'_z \;
{V\over (2\pi)^3} \, d^3P'
\,. 
\label{drho}
\end{eqnarray}
Therefore
\begin{eqnarray}
d\sigma &=& V\, {|S_{fi}|^2\over T} \; {L_xL_z\over (2\pi)^2} 
\, dp'_x \, dp'_z \;
{V\over (2\pi)^3} \, d^3P' \nonumber\\*
&=& {1\over 64\pi^2} \delta^3 \omit y (P+k-P'-p') \;
{E_{n'}+m \over \omega {\cal E}{\cal E}' E_{n'}} \;
\Big| {\mathscr M}_{fi} \Big|^2 {L_xL_z\over V} 
\, dp'_x \, dp'_z \; d^3P' \,.
\label{dsigma}
\end{eqnarray}
The square of the matrix element is
\begin{eqnarray}
\Big| {\mathscr M}_{fi} \Big|^2 = 2G_\beta^2 \ell^{\mu\nu} H_{\mu\nu} \,,
\end{eqnarray}
where $H_{\mu\nu}$ is the hadronic part and $\ell^{\mu\nu}$ the
leptonic part, whose calculation we outline now.

For the hadronic part, we can use the usual Dirac spinors because of
our assumption that the magnetic field is much smaller than $m_p^2/e$.
We will work in the rest frame of the neutron.  Due to the presence of
the background magnetic field, the neutrons may be totally or
partially polarized.  We define the quantity 
\begin{eqnarray}
S \equiv {N_n^{(+)} - N_n^{(-)} \over N_n^{(+)} + N_n^{(-)}} \,,
\end{eqnarray}
where $N_n^{(\pm)}$ denote the number of neutrons parallel and
antiparallel to the magnetic field.  Then
\begin{eqnarray}
H_{\mu\nu} = \frac12 (1+S) H_{\mu\nu}^{(+)} + 
\frac12 (1-S) H_{\mu\nu}^{(-)} \,,
\end{eqnarray}
where $H_{\mu\nu}^{(\pm)}$ denotes the contribution calculated with
spin-up and spin-down neutrons respectively.  Either of these
contributions can be calculated by using the spin projection operator,
which is $\frac12(1\pm\gamma_5\gamma_3)$ for up and down spins.  A
straight forward calculation then yields
\begin{eqnarray}
H_{\mu\nu} &=& 2(g_V^2 + g_A^2) (P_\mu P'_\nu + P_\nu P'_\mu -
g_{\mu\nu} P \cdot P') \nonumber\\*
&& + 2 (g_V^2-g_A^2) m_n m_p g_{\mu\nu} + 4i g_V
g_A \varepsilon_{\mu\nu\lambda\rho} P^\lambda P'^\rho \nonumber\\*
&-& S \Big[ 
4 g_Vg_A m_n (P'_\mu g_{3\nu} + P'_\nu g_{3\mu} - P'_3
g_{\mu\nu})
+ 2 i \varepsilon_{\mu\nu3\alpha} R^\alpha \Big]  
\,,
\end{eqnarray}
where we have introduced the shorthand
\begin{eqnarray}
R^\alpha = (g_V^2 + g_A^2) m_n P'^\alpha 
- (g_V^2-g_A^2) m_p P^\alpha \,.
\end{eqnarray}
We have omitted some terms in the expression for $H_{\mu\nu}$ that
involve spatial components of the neutron momentum, with the
anticipation that we will perform the calculation in the neutron rest
frame. 

In the leptonic part $\ell^{\mu\nu}$, we should use the magnetic
spinors given in Sec.~\ref{so}.  This gives
\begin{eqnarray}
\ell^{\mu\nu} = \int dy \int dy_\star \; e^{iq_y(y_\star-y)} \; {\rm Tr}
\Big[ P_U (y, y_\star, n', \vec p' \omit y) \gamma^\mu \rlap/k
\gamma^\nu L \Big] \,,
\end{eqnarray}
where $P_U$ denotes the spinor sum for the electrons, given in Eq.\
(\ref{PU}).  We now have to perform the integrations over $y$ and
$y_\star$.  Each of these variables should be integrated in the range
$-\frac12L_y$ to $+\frac12L_y$.  However, since we will take the
infinite volume limit at the end as usual, we let $L_y\to\infty$ and
use the result~\cite{GradRyzh}
\begin{eqnarray}
\int_{-\infty}^{+\infty} dy \; e^{ixy} I_n(y) 
= i^n \; \sqrt{2\pi} \; I_n (x) 
\,.
\end{eqnarray}
This gives
\begin{eqnarray}
\ell^{\mu\nu} = {2\pi\over eB} \; {1 \over (E_{n'}+m)} (\Lambda^\mu
k^\nu + \Lambda^\nu k^\mu - k \cdot \Lambda g^{\mu\nu} - i
\varepsilon^{\mu\nu\alpha\beta} \Lambda_\alpha k_\beta) \,,
\end{eqnarray}
where
\begin{eqnarray}
\Lambda^\alpha &=& \left[I_{n'-1} 
\left({q_y \over \sqrt{eB}} \right) \right]^2 
(p'^\alpha_\parallel - \widetilde p'^\alpha_\parallel) 
+ \left[ I_{n'} \left({q_y \over \sqrt{eB}} \right) \right]^2 
(p'^\alpha_\parallel + \widetilde p'^\alpha_\parallel) \nonumber\\* && 
- 2 \sqrt{2n'eB} g_2^\alpha I_{n'}
\left({q_y \over \sqrt{eB}} \right) I_{n'-1}
\left({q_y \over \sqrt{eB}} \right) \,.
\label{Lambda}
\end{eqnarray}
Thus,
\begin{eqnarray}
\Big| {\mathscr M}_{fi} \Big|^2 &=& 8G_\beta^2 \times {2\pi\over eB} \;
{1 \over (E_{n'}+m)}  \Bigg[ (g_V^2+g_A^2) (P\cdot
\Lambda \;
P'\cdot k + P'\cdot \Lambda \; P\cdot k)\nonumber\\* 
&&  
- (g_V^2-g_A^2) m_nm_p k \cdot \Lambda - 2 g_Vg_A (P\cdot \Lambda \;
P'\cdot k - P'\cdot \Lambda \; P\cdot k) \nonumber\\* 
&& + S \Big( 2 g_Vg_A m_n
(P' \cdot \Lambda k_z + P' \cdot k \Lambda_z) - \Lambda_z k \cdot R +
k_z \Lambda \cdot R \Big) 
\Bigg]  \,. 
\end{eqnarray}

We now choose the axes such that the 3-momentum of the incoming
neutrino is in the $x$-$z$ plane.  We will also assume that $|\vec P'|
\ll m_p$ for the range of energies of interest to us. In that case, it
is easy to see that the terms involving $\sqrt{2n'eB}$ drop out, and
we obtain
\begin{eqnarray}
\Big| {\mathscr M}_{fi} \Big|^2 = 8G_\beta^2 \times {2\pi\over eB} \;
{m_nm_p \over E_{n'}+m} &\times& \Big[ (g_V^2+3g_A^2) \omega \Lambda_0
+ (g_V^2-g_A^2) k_z \Lambda_z \nonumber\\*
&& + 2g_A S \Big( (g_V-g_A) \omega \Lambda_z + (g_V+g_A) k_z\Lambda_0 
\Big) \Big] \,. 
\end{eqnarray}

We now put this expression into Eq.\ (\ref{dsigma}) and calculate the
total cross section by performing the integrations over different
final state momenta appearing in that formula.  First we integrate
over $P'_x$ and $P'_z$.  These appear only in the momentum conserving
$\delta$-function.  Integration over them therefore just gets rid of
the corresponding $\delta$-functions.  For the integration over
$p_x'$, we refer to Eq.\ (\ref{xi}). Since the center of the
oscillator has to lie between $-\frac12 L_y$ and $\frac12 L_y$, we
conclude that $-\frac12 L_yeB\leq p'_x\leq\frac12 L_yeB$. Thus the
integration over $p_x'$ gives a factor $L_yeB$.

Putting back into Eq.\
(\ref{dsigma}) and using $V=L_xL_yL_z$, we obtain
\begin{eqnarray}
d\sigma 
= {G_\beta^2\over 4\pi} {\delta (Q+\omega-E_{n'})
 \over \omega E_{n'}} &\times& 
\Big[ (g_V^2+3g_A^2) \omega \Lambda_0
+ (g_V^2-g_A^2) k_z \Lambda_z \nonumber\\* 
&& + 2g_AS \Big( (g_V-g_A) \omega \Lambda_z + (g_V+g_A) k_z\Lambda_0 
\Big) \Big] \; dP'_y dp'_z \,,
\label{dsigma2}
\end{eqnarray}
where $Q$ is the neutron-proton mass difference, $m_n-m_p$.

We next perform the integration over $P'_y$. In the integrand, it
occurs only as the argument of the functions $I_n$ and $I_{n-1}$.
The functions $I_n$ are orthogonal in the sense that
\begin{eqnarray}
\int_{-\infty}^{+\infty} da \; I_n(a) I_{n'}(a) 
= \sqrt{eB} \; \delta_{nn'} \,.
\end{eqnarray}
This property can be used to perform the integration over $P'_y$.  We
have already remarked that the term proportional to $\sqrt{2n'eB}$ in
Eq.\ (\ref{Lambda}) does not contribute.  {}From other two terms, we
obtain
\begin{eqnarray}
\int dP'_y \, \Lambda^\alpha &=& eB \Big[ (p'^\alpha_\parallel -
\widetilde p\,'^\alpha_\parallel) (1 - \delta_{n',0}) 
+ (p'^\alpha_\parallel + \widetilde p\,'^\alpha_\parallel) \Big]
\nonumber\\* 
&=& eB \Big[ g_{n'} p'^\alpha_\parallel + \delta_{n',0}
\widetilde p\,'^\alpha_\parallel \Big] \,,
\label{intdP'y}
\end{eqnarray}
where
\begin{eqnarray}
g_{n'} = 2 - \delta_{n',0}
\end{eqnarray}
gives the degeneracy of the Landau level.  Notice the appearance of
the Kronecker delta, $\delta_{n',0}$, in the expression of Eq.\
(\ref{intdP'y}).  The reason for this is that, while two terms of Eq.\
(\ref{Lambda}) contribute in the integral for $n'\neq0$, only one of
them contributes for $n'=0$ since $I_{-1}=0$.

The final integration is over $p'_z$.  Writing the argument of the
remaining $\delta$-function in terms of $p'_z$, we find that the zeros
occur when
\begin{eqnarray}
p'_z = p'_\pm \equiv \pm \sqrt{(Q+\omega)^2 - m^2 - 2n'eB} \,.
\end{eqnarray}
Therefore, 
\begin{eqnarray}
\delta (Q+\omega-E_{n'}) = {Q+\omega \over \sqrt{(Q+\omega)^2 - m^2 -
2n'eB}} \Big( \delta (p'_z - p'_+) + \delta (p'_z - p'_-) \Big) \,.
\end{eqnarray}
In the integration, the terms proportional to $p'_z$ in the integrand
receive equal and opposite contributions from the two $\delta$
functions and cancel.  For the other terms, independent of $p'_z$,
both the contributions are equal.  So we obtain
\begin{eqnarray}
\sigma_{n'} 
= {eBG_\beta^2\over 2\pi} && 
\bigg[ g_{n'} \Big\{ (g_V^2+3g_A^2)  + 2g_AS(g_V+g_A) \cos\theta \Big\}
\nonumber\\* 
&& +  \delta_{n',0} \Big\{ (g_V^2-g_A^2) \cos\theta 
+ 2g_AS (g_V-g_A) \Big\} \bigg] 
 {Q+\omega \over \sqrt{(Q+\omega)^2 - m^2 - 2n'eB}} \,, \quad
\label{sigman'}
\end{eqnarray}
where we have defined the direction of the incoming neutrino by the
angle $\theta$, with
\begin{eqnarray}
k_z = \omega \cos\theta \,.
\end{eqnarray}

In Eq.\ (\ref{sigman'}), we have denoted the cross section by
$\sigma_{n'}$ because the electron ends up in a specific Landau level
$n'$.  The total cross section is then given as a sum over all
possible values of $n'$, i.e.,
\begin{eqnarray}
\sigma = \sum_{n'=0}^{n'_{\rm max}} \sigma_{n'} 
&=& {eBG_\beta^2\over
2\pi} \sum_{n'=0}^{n'_{\rm
max}} 
\bigg[ g_{n'} \Big\{ (g_V^2+3g_A^2)  + 2g_AS(g_V+g_A) \cos\theta \Big\}
\nonumber\\* 
&& +  \delta_{n',0} \Big\{ (g_V^2-g_A^2) \cos\theta 
+ 2g_AS (g_V-g_A) \Big\} \bigg] 
 {Q+\omega \over \sqrt{(Q+\omega)^2 - m^2 - 2n'eB}}  \,. \qquad
\label{sigma}
\end{eqnarray}
The possible allowed Landau level has a maximum, $n'_{\rm max}$, which
is given by the fact that the quantity under the square root sign in
the denominator of Eq.\ (\ref{sigma}) must be
non-negative, i.e.,
\begin{eqnarray}
n'_{\rm max} = {\rm int} \left\{ {1\over 2eB} \Big[(Q+\omega)^2-m^2 
\Big] \right\} \,.
\label{n'max}
\end{eqnarray}

Eq.\ (\ref{sigma}) gives our result for the cross section of the
inverse beta decay process.  Some properties of this formula are worth
noting. 

For unpolarized neutrons, $S=0$, the cross section for $n'\neq0$ does
not depend on the direction of the incoming neutrino.  The same is not
true if the electron ends up in the lowest Landau level.  The cross
section will be asymmetric in this case.

All terms in the cross section which depend on $S$ have a common
factor $g_A$.  The reason is that, if $g_A$ were equal to zero, the
interaction in the hadronic sector would have been spin-independent.

If the final electron is in the lowest Landau level and the initial
neutrino momentum is antiparallel to the magnetic field, Eq.\
(\ref{sigma}) shows that
\begin{eqnarray}
\sigma_0 = {eBG_\beta^2\over 2\pi} 
\bigg[  4g_A^2 (1-S) \bigg] 
 {Q+\omega \over \sqrt{(Q+\omega)^2 - m^2}} \,. 
\end{eqnarray}
Note that the vector coupling $g_V$ does not contribute to the cross
section in this limit.  This can be understood easily.  The neutrino
spin is along the $+z$ direction whereas the electron spin in the
lowest Landau level must be in the $-z$ direction.  Thus there is a
spin-flip in the leptonic sector.  Conservation of angular momentum
then implies that there must be a spin-flip in the hadronic sector as
well.  In the non-relativistic limit for hadrons that we have
employed, this can occur only through the axial coupling.

If further we consider totally polarized neutrons, i.e., $S=1$, we see
that $\sigma_0$ vanishes.  Again, this is a direct consequence of
angular momentum conservation.  Since both initial particles have spin
up, angular momentum conservation requires both final particles in
spin up states as well.  But the spin-up state is not available for
the electron in the lowest Landau level.

\begin{figure}[tb]
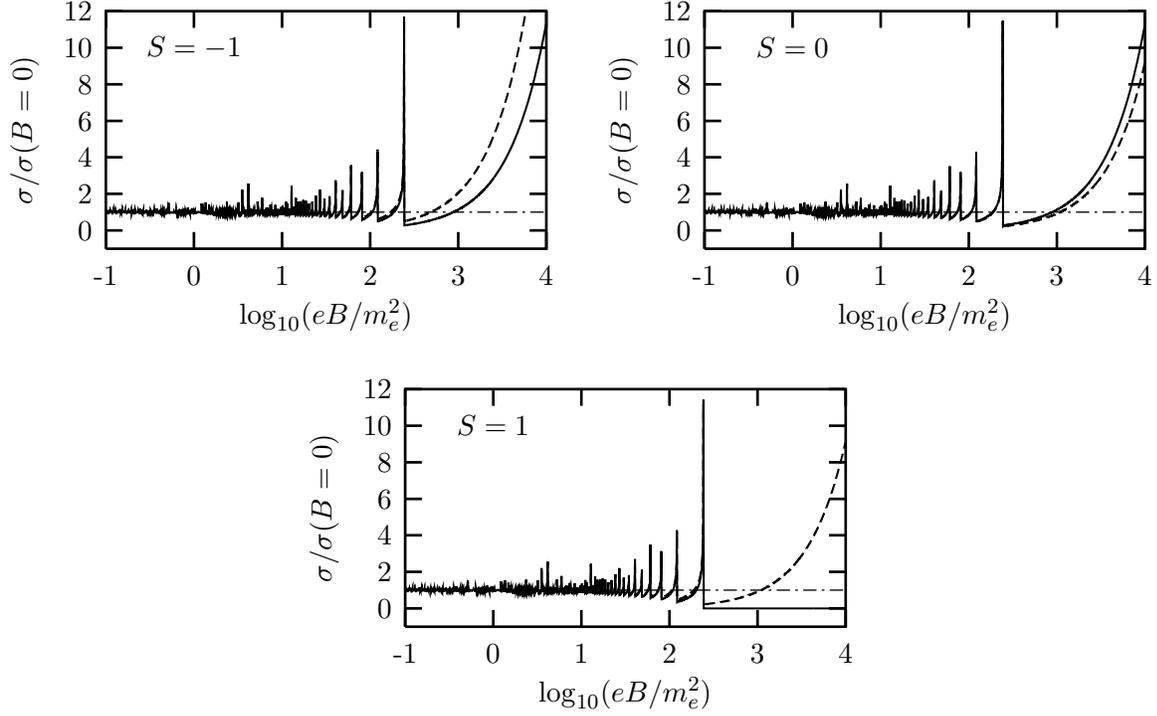

\begin{center}\input{sigma_Sneg.psl} \hfil 
\input{sigma_S0.psl} \\  \bigskip
\input{sigma_Spos.psl}
\end{center}
\caption[]{\small\sf Enhancement of cross section in a magnetic field
for an initial neutrino energy of 10\,MeV.  Different panels show the
results for different net polarizations of the neutrons.  The solid
and the dashed lines correspond to the initial neutrino
momentum parallel and antiparallel to the magnetic field.  Each curve
has been normalized by the cross section in field-free case for the
same values of $S$ and $\cos\theta$.  The horizontal dashed lines
represent unity in the vertical scale.}\label{f:sigma}
\end{figure}
It is instructive to check that the result obtained in Eq.\
(\ref{sigma}) reduces to the known result for the field-free case. The
contribution specific to the zeroth Landau level vanishes in the limit
$B\to0$ owing to the overall factor of $eB$.  The other terms also
have the factor $eB$, but in this case we also need to sum over
infinitely many states. This gives
\begin{eqnarray}
\sigma &=& {eBG_\beta^2\over \pi} 
\Big[ (g_V^2+3g_A^2) + 2g_A S (g_V+g_A) \cos\theta \Big] \nonumber\\*
&& \times \left( \sum_{n'=0}^{n'_{\rm max}} {Q+\omega \over
\sqrt{(Q+\omega)^2 - m^2 - 2n'eB}} - {Q+\omega \over
2\sqrt{(Q+\omega)^2 - m^2}} \right)\,.
\end{eqnarray}
For $B\to 0$, the last
term vanishes, and we can identify $n'_{\rm max}$ as the integer for
which the denominator of the summand vanishes. Thus we obtain
\begin{eqnarray}
\sigma &\longrightarrow& {eBG_\beta^2\over \pi} 
\Big[ (g_V^2+3g_A^2) + 2g_A S (g_V+g_A) \cos\theta \Big]
\int_0^{n'_{\rm max}} dn'\; {Q+\omega \over
\sqrt{(Q+\omega)^2 - m^2 - 2n'eB}} \nonumber\\*
&=& {G_\beta^2\over \pi}  
\Big[ (g_V^2+3g_A^2) + 2g_A S (g_V+g_A) \cos\theta \Big]
(Q+\omega) \sqrt{(Q+\omega)^2 - m^2} \,,
\end{eqnarray}
which is the correct result in the field-free case.

In Fig.~\ref{f:sigma}, we have plotted the ratio of the cross section
to its corresponding value at $B=0$ as a function of the magnetic
field.  The plots have been done for unpolarized ($S=0$) as well as
totally polarized neutrons along ($S=1$) and opposite ($S=-1$) to the
magnetic field, with the initial neutrino momentum parallel and
antiparallel to the magnetic field.  For $S=-1$, we find that
neutrinos parallel to the magnetic field have a smaller cross section
than those antiparallel to the field, and the difference is pronounced
for large fields.  For $S=0$, the situation is just reversed.  For
$S=1$, if the magnetic field is high enough so that $n'_{\rm max}=0$,
we see that the cross section vanishes for neutrino momentum
antiparallel to the field.  The reason for this has already been
discussed.

\section{Consequences of neutrino energy spread}\label{es}
The enhancement factor in Fig.~\ref{f:sigma} shows some spikes.  They
appear at values of the magnetic field for which the denominator of
Eq.\ (\ref{sigma}) vanishes for some $n'$.  For field values larger
than this, that particular Landau level does not contribute to the
cross section.  To the right of the final spike that appears in the
figure, only the zeroth Landau level contributes.  In other words, the
final electron can go only to the lowest Landau level for such high
values of the magnetic field.  The exact value of $B$ for which this
occurs depends of course on the energy of the initial neutrino.

We need to make an important point about these spikes.  Each spike in
fact go all the way up to infinity.  The finite height of a spike in
the figure is an artifact of the finite step size taken in plotting
it.

In reality, of course, a cross section cannot be infinite. In the
present case, this is related to the fact that the initial neutrinos
cannot be exactly monochromatic due to the uncertainty relation.
There must be a spread in energy, which can be represented by a
probability distribution $\Phi(\omega)$, defined by
\begin{eqnarray}
\int d\omega \; \Phi(\omega) = 1 \,.
\end{eqnarray}
In that case, the cross section in a real experiment should be written
in the form
\begin{eqnarray}
\sigma = \int d\omega \; \Phi(\omega) \sigma(\omega) \,,
\label{sigmaint}
\end{eqnarray}
where $\sigma(\omega)$ is the expression derived in Eq.\
(\ref{sigma}) for a single value of energy.

\begin{figure}[tbp]
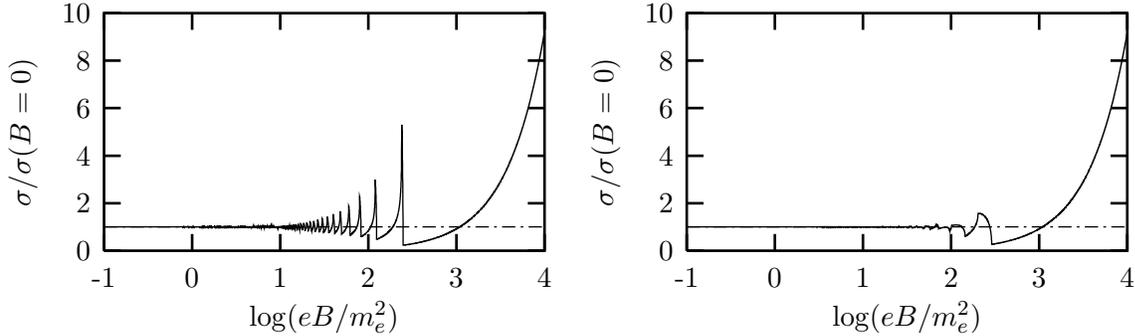

\centerline{\input{smooth1.psl} 
\input{smooth2.psl} }
\caption[]{\small\sf Cross section for unpolarized neutrons as a
function of the magnetic 
field for a flat energy distribution, normalized to the cross section
in the field-free case. The initial neutrino momentum is along the
magnetic field, and energy is 10\,MeV.  The energy spread
$\omega_2-\omega_1$ is $0.2$\,MeV for the left panel and 2\,MeV for
the right panel.}\label{f:smooth}
\end{figure}%
As an illustration, we consider the case of unpolarized neutrons
($S=0$), and take a flat probability distribution of initial neutrino
energy, viz.,
\begin{eqnarray}
\Phi(\omega) = \cases{{\displaystyle 1\over \displaystyle
\omega_2 - \omega_1} & if $\omega_1 \leq \omega \leq \omega_2$, \cr \cr
0 & otherwise.}
\label{Edistrn}
\end{eqnarray}
Then the integration of Eq.\ (\ref{sigmaint}) gives
\begin{eqnarray}
\sigma &=& {eBG_\beta^2\over 2\pi (\omega_2 - \omega_1)} \Big[
F(\omega_2) - F(\omega_1) \Big] \,,
\end{eqnarray}
where
\begin{eqnarray}
F(\omega) = \sum_{n'=0}^{n'_{\rm max}}
\Bigg[ g_{n'} (g_V^2+3g_A^2) 
+ \delta_{n',0} (g_V^2-g_A^2) \cos\theta 
\Bigg] 
\times{\sqrt{(Q+\omega)^2 - m^2 - 2n'eB}} \,,
\end{eqnarray}
with $n'_{\rm max}$ determined by Eq.\ (\ref{n'max}).
Fig.~\ref{f:smooth} shows the variation of this quantity with the
magnetic field for $\cos\theta=1$.  In this figure, we normalize the
cross section by $B=0$ cross section with the energy distribution of
Eq.\ (\ref{Edistrn}), which is
\begin{eqnarray}
\sigma (B=0) = {G_\beta^2 (g_V^2+3g_A^2) \over 3\pi (\omega_2-\omega_1)} 
\bigg( \Big[(Q+\omega_2)^2 - m^2 \Big]^{3/2}
- \Big[(Q+\omega_1)^2 - m^2 \Big]^{3/2} \bigg) \,.
\end{eqnarray}
Keeping the central value of neutrino energy as 10 MeV as before, we
have drawn these plots for two different values of the spread, as
mentioned in the caption.  For the smaller value of the spread in
particular, the graph looks very similar to that drawn in
Fig.~\ref{f:sigma}, but the difference is that now the height of the
spikes denote the actual enhancement, and is not an artifact of the
plotting procedure.  For the higher value of the energy spread, we see
that the spikes have smoothened out.

\section{Comments}\label{co}
The calculation of the cross section for inverse beta decay process
has been performed earlier by several authors \cite{Roulet:1997sw,
Lai:1998sz}.  They assumed that the matrix element remains unaffected
by the magnetic field, only the modified phase space integral makes
the difference in the cross section. The results they obtained is the
same as the term proportional to $g_V^2+3g_A^2$ that we obtained.

We have used exact spinor solutions in a uniform magnetic field to
calculate the cross section of the inverse beta decay process for
arbitrary neutron polarization.  We find that, even for unpolarized
neutrons, the cross section depends on the direction of the neutrino
momentum.  This asymmetry is not surprising since the background
magnetic field makes it an anisotropic problem.  A similar asymmetry
has been noted for URCA processes \cite{Baiko:1998jq}, where the
neutrino is in the final state.  The anisotropy for the $S=0$ case
comes only from the $n'=0$ contribution
\cite{Gvozdev:1999md}. However, for $n'\neq 0$, there is a
cancellation between the two possible states in a Landau level which
washes out all angular dependence in these levels, provided the
neutrons are unpolarized.  The asymmetry in cross section will
therefore come only from the $n'=0$ state and its amount will depend
on the relative contribution of this state to the total cross
section. If the magnetic field is so high that only the $n'=0$ state
can be obtained for the electron, the asymmetry will be large, about
18\%. For smaller and smaller magnetic fields, the asymmetry decreases
with new Landau levels contributing.  For polarized neutrons, however,
there is an asymmetry even in the field-free case.  In presence of a
magnetic field, the asymmetry will in general depend on the magnetic
field, as it appears from various plots in Fig.~\ref{f:sigma}.

This fact can have far reaching consequences for neutrino emission
from a proto-neutron star. It has been discussed in the literature
that the presence of asymmetric magnetic fields can cause asymmetric
neutrino emission from a proto-neutron star \cite{Bisno}. However, our
calculations show that even with a uniform magnetic field, neutrino
emission would be asymmetric because of the $\cos\theta$-dependent
terms in the cross section.

\paragraph*{Acknowledgements~: } 
We thank Debades Bandopadhyay, A.~Y. Parkhomenko and Esteban Roulet
for illuminating discussions and for pointing out some important
papers in the field.  We also thank Duane Dicus and Huaiyu Duan for
pointing out some mistakes in an earlier version of our work.

\end{document}